\documentclass[12pt, preprint]{emulateapj}
 \slugcomment{Accepted to ApJ}
\usepackage{apjfonts}
 \newcommand{\Jnic}{$J_{110}$}
 \newcommand{\Hnic}{$H_{160}$}
 \newcommand{\Iacs}{$I_{814}$}
 \newcommand{\zacs}{$z_{850}$}
 \newcommand{\ip}{$i'$}
 \newcommand{\zp}{$z'$}

 \newcommand{\jdtwo}{JD2325+1433}
 \newcommand{\bgoods}{$B_{435}$}
 \newcommand{\vgoods}{$V_{606}$}
 \newcommand{\igoods}{$i_{775}$}

\begin{document}
\title{Expanding the search for galaxies at  $\lowercase{z} \sim 7-10$ with new NICMOS Parallel Fields \altaffilmark{1}}
 
 \author{Alaina L. Henry\altaffilmark{2},   Brian Siana\altaffilmark{3},   Matthew A. Malkan\altaffilmark{2}, 
Matthew L.~N. Ashby\altaffilmark{4}, Carrie R. Bridge\altaffilmark{5},   Ranga-Ram Chary\altaffilmark{5}, James W. Colbert\altaffilmark{5},    Mauro Giavalisco\altaffilmark{6},  Harry I. Teplitz\altaffilmark{5},  \& Patrick J. McCarthy \altaffilmark{7}}
 
 \altaffiltext{1}{
 This work is  based in part on observations made with the NASA/ESA {\it Hubble Space Telescope}, obtained from the Space Telescope Science Institute, which is operated by the Association of Universities for Research in Astronomy Inc., under NASA contract NAS 5-26555.  These observations are associated with proposals 10872, 11236, and 11188.   This work is also based in part on observations made with the Spitzer Space Telescope, which is operated by the Jet Propulsion Laboratory, California Institute of Technology under a contract with NASA. Support for this work was provided by NASA through an award issued by JPL/Caltech. }
\altaffiltext{2}{Department of Physics and Astronomy, Box 951547, UCLA, Los Angeles, CA 90095, USA; ahenry@astro.ucla.edu                   }
  \altaffiltext{3}{California Institute of Technology, MS 105-24, Pasadena, CA 91125}
  \altaffiltext{4}{Harvard-Smithsonian Center for Astrophysics; 60 Garden Street, MS-66, Cambridge, MA, 02138}
 \altaffiltext{5}{$Spitzer$ Science Center, California Institute of Technology, 220-6, Pasadena, CA, 91125, USA}
 \altaffiltext{6}{Astronomy Department, University of Massachusetts, Amherst, MA 01003}
\altaffiltext{7}{Observatories of the Carnegie Institute of Washington, Santa Barbara Street, Pasadena, CA 91101}

 \begin{abstract}
 We have carried out a search for galaxies at $ z \sim 7-10 $  in $\sim$ 14.4 arcmin$^2$ of new NICMOS 
 parallel imaging taken in the Great Observatories Origins Deep Survey (GOODS, 5.9 arcmin$^2$),  the Cosmic Origins Survey (COSMOS,  7.2 arcmin$^2$), and SSA22 (1.3 arcmin$^2$).  These images reach 5 $\sigma$ sensitivities of \Jnic\ = 26.0-27.5 (AB), and combined they increase the amount of deep near-infrared data by more than 60\% in fields where the investment in deep optical data has already been made. 
  We find no $z>7$ candidates in our survey area,  consistent with the Bouwens et al.\ (2008) measurements at $z\sim7$ and 9 (over 23 arcmin$^2$), which predict 0.7 galaxies at $z\sim7$ and $<0.03$ galaxies at $z\sim9$.    We estimate that 10-20 \% of $z>7$ galaxies are missed by this survey, due to incompleteness from foreground contamination by faint sources.  For the case of luminosity evolution, assuming a Schecter parameterization with a typical $\phi^* = 10^{-3} ~{\rm Mpc}^{-3}$, we find $M^* > -20.0$ for $z\sim7$ and $M^* > -20.7$ for $z\sim9$ (68\% confidence).     This suggests that the downward luminosity evolution of LBGs continues to $z\sim7$, although our result is marginally consistent with the $z\sim6$ LF of Bouwens et al.\ (2006, 2007).     In addition we present newly-acquired deep MMT/Megacam imaging of the $z\sim9$ candidate \jdtwo, first presented in Henry et al.\ (2008).  The resulting weak but significant detection at  \ip\   indicates that this galaxy is most likely an interloper at $z\sim2.7$.
  \end{abstract}

\section{Introduction}
Populations of Lyman break galaxies (LBGs) have now been identified up to $z\sim6$, when   
the universe was less than 1 Gyr old.  Observations now point to earlier times as an important period in the evolution of galaxies.   First, some galaxies at $z\sim6$ have  well established stellar populations, with ages $\sim100$ Myrs and masses $\ga 10^{10}$ M$_{\sun}$ (\citealt{Eyles05}, 2007; \citealt{Yan06}; \citealt{Verma}; \citealt{Stark07}), requiring significant star 
formation at $z>7$.   Second, these  ``first galaxies''  likely played an essential role in the reionization of the intergalactic medium, which occurred sometime between $z\sim7$ and 14 (from the {\it Wilkinson Microwave Anisotropy Probe}, WMAP; \citealt{Dunkley}).    

Observations of these $z\ga7$ galaxies are crucial; however, the search has been significantly more difficult than surveys for LBGs at $z\sim3-6$.    At  $z\sim7-8$, the Lyman break passes into the $z$-band, and galaxies must be identified with near-infrared imaging, where sensitivity and area are limited.   To make matters more challenging, evolution of the  UV luminosity function shows declining numbers of luminous LBGs with increasing redshift, over the period of $z\sim3-6$ \citep{Bouwens07}.   Regardless of whether this trend continues to $z\sim7$, the low density of luminous LBGs at $z\sim6$ (a few hundred degree$^{-2}$ to \zacs\ = 26)  means that both wide area and sensitivity are necessary to continue the search to $z>7$.

Progress in this search for high$-z$ LBGs has been made on three fronts.   First, wide-area surveys have probed the bright end of the luminosity function (LF).  \cite{Mannucci} used the VLT/ISAAC NIR data in GOODS South to search 130 arcmin$^2$ to $J\sim25.5$, and  \cite{Stanway08b} searched eleven independent sight lines covering 360 arcmin$^{2}$ to $J_{AB} = 24-25$. Both teams find only a few marginal candidates which they interpret as probable interlopers.  Their limits are roughly consistent with extrapolation of the $z\sim6$ LF, although \cite{Mannucci} report a slight decline to $z\sim7$.  At higher redshifts, we have searched 135 arcmin$^2$ of deep \Jnic\ and \Hnic\  parallel images for galaxies at $z\sim8-10$, uncovering one $z\sim9$ candidate (\jdtwo; \citealt{Henry08}).

A second approach has been to use strong  gravitational lensing to probe the fainter luminosities, where the  volume density of  
$z>7$ galaxies should be higher.  Several candidates have been found by this technique (\citealt{Bradley}; \citealt{richard}, 2008).  However, in an independent analysis of the \cite{richard08}  data, \cite{Bouwens08b}  suggest that most of these galaxies are either spurious detections, or they fail to meet the $z>7$ selection criteria.  This disagreement is indicative of the challenge posed by the search for these extremely faint galaxies.   To make progress, very deep observations are needed in both the optical and near-infrared. 

\begin{figure*}
\plottwo{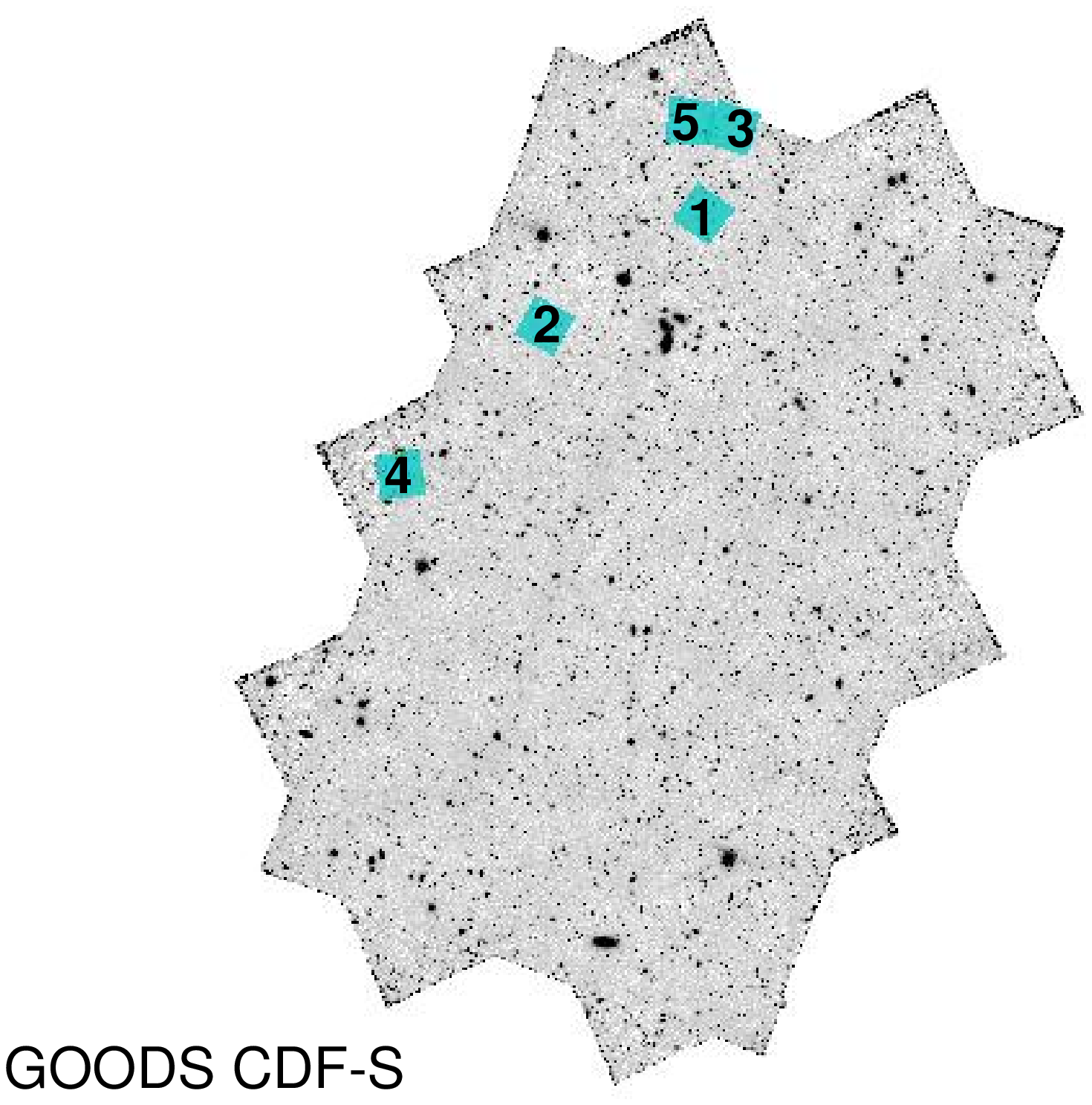} {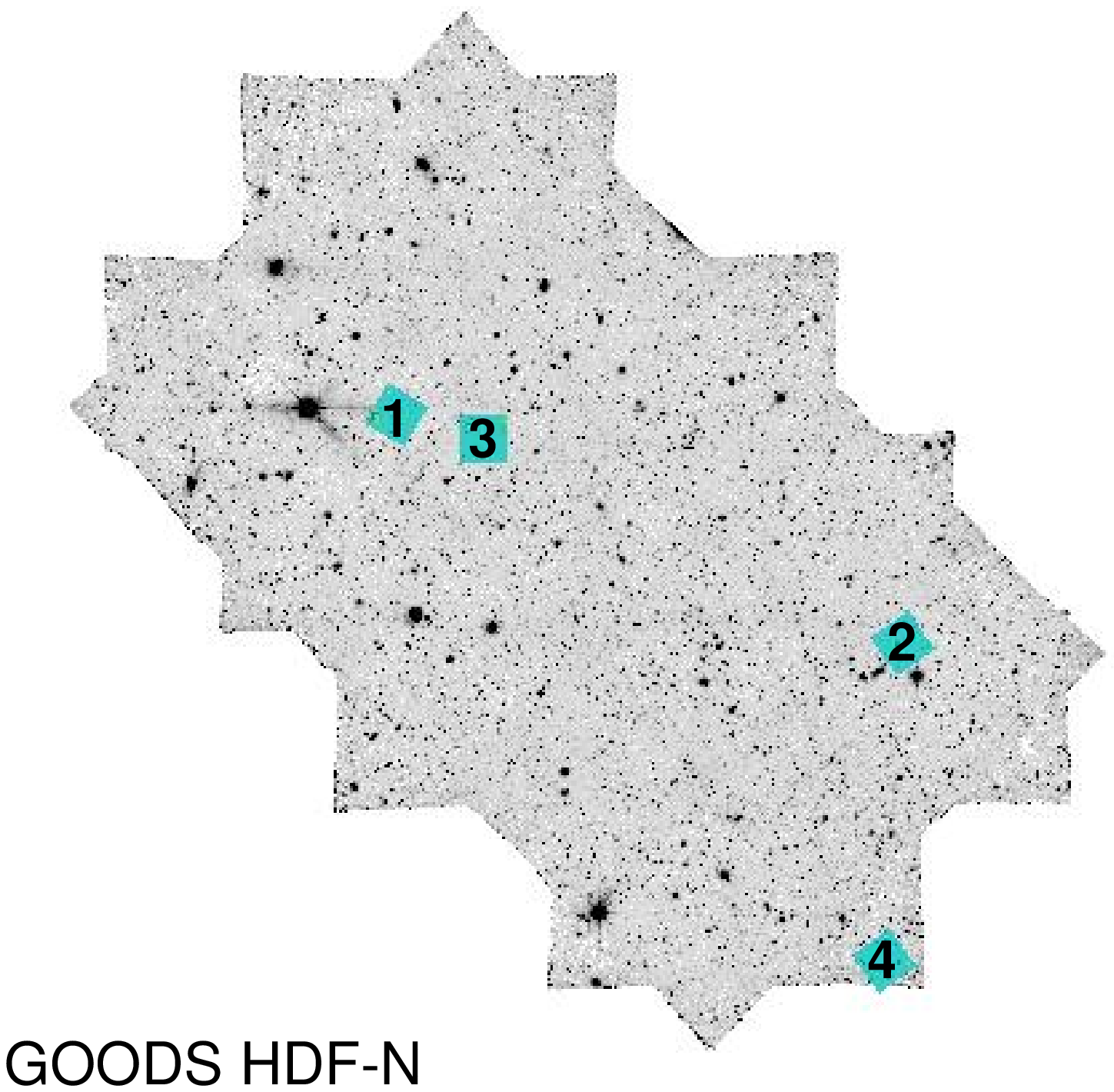} 
\caption{Cyan squares show the new NICMOS parallel fields in GOODS which we use to search for $z>7$ galaxies.  
Comparison to Figure 1 from Bouwens et al. 2008 shows that CDFS-1, -2, and -3 are also used as part of their survey.  Coordinates of these fields are listed in Table \ref{obstable}.  }
\label{goods_footprint}
\end{figure*}

This challenge is mitigated by the use of deep NICMOS imaging in GOODS,  including the Ultra Deep Field (UDF), the Hubble Deep Field North (HDFN), and various parallel exposures (\citealt{Bouwens04}; 2005, 2008;  \citealt{Labbe}; \citealt{Oesch}).  Although only eight candidates are found in these $\sim 23$ arcmin$^2$, and none are spectroscopically confirmed, \cite{Bouwens08a} report a luminosity function with a bright end that continues to evolve in the same manner as those at $z\sim3-6$.   While uncertain, these data suggest that the density of the most luminous $z\sim7$ galaxies is even smaller than at $z\sim6$.

\begin{figure}
\plotone{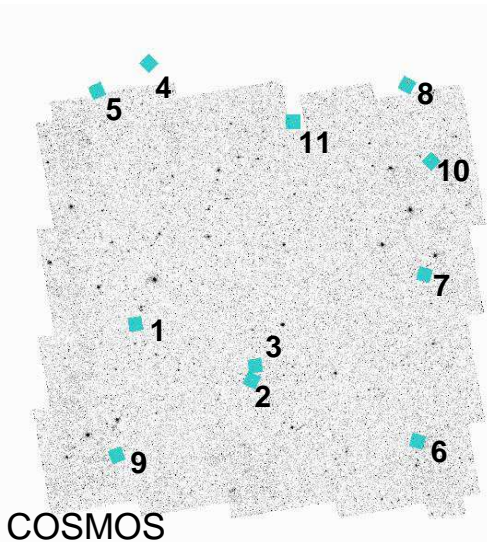}
\caption{Cyan squares show the new NICMOS parallel fields in COSMOS, overlaid on the ACS \Iacs\ mosaic.   For visualization, squares are enlarged from the actual NICMOS footprint by a factor of nine in area.  While fields 4, 5, 8, and 11 are outside the ACS \Iacs\ mosaic, they are within the Subaru/SuprimeCam \zp and \ip\ images.  Coordinates of these fields are listed in Table \ref{obstable}.     }

\label{cosmos_footprint}
\end{figure}

Because only eight of these galaxies have been found, expanding the most sensitive combined infrared and optical coverage to identify even one additional $z-$dropout LBG would be a significant contribution. 
 Accordingly,  we have obtained $\sim 14.4$ arcmin$^2$ of  coordinated NICMOS parallel observations in \Jnic\ and \Hnic, taken in the GOODS fields (\citealt{Giavalisco}), COSMOS (\citealt{cosmos}; \citealt{Koekemoer}) and SSA22  \citep{Steidel98}.  The GOODS and COSMOS images reach 5 $\sigma$ = 26.0-26.7 in \Jnic\ (0\farcs6 diameter aperture)-- 1-2 magnitudes deeper than the wide area ground based searches carried out by \cite{Mannucci} and \cite{Stanway08b}.   The two parallel fields in SSA22 are significantly deeper, reaching 5 $\sigma = 27.5$ and 27.0 in \Jnic.   Although most of this area is less sensitive than the  UDF and HDFN,  four out  of eight $z\sim7$ candidates  in Bouwens et al. (2008;  $\sim 23$ arcmin$^2$) are bright enough to be detected in the deepest of these new GOODS and COSMOS images, and most are bright enough to be detected in the SSA22 fields.   In addition to this search, we have carried out deep follow-up optical imaging of \jdtwo, the $z\sim9$ candidate presented in \cite{Henry08}.

 In \S \ref{data} we describe the data reduction and photometry, as well as  an overview of the public data products that we use.    In \S \ref{selection} we describe the selection of $z>7$ candidates and the criteria which we use to discriminate against interlopers.  In \S \ref{discussion} we derive a new upper limit on the volume density of $z\sim7-8$ galaxies, and discuss implications for the reionization of hydrogen in the intergalactic medium.  Finally, in \S \ref{jd2_followup} we present new observations of the $z\sim9$ candidate mentioned above, which suggest that it is an intermediate redshift interloper.        We use $H_0 = 70 ~{\rm   km s^{-1}~ Mpc^{-1}}$, $\Omega_{\Lambda} = 0.7$, $\Omega_M = 0.3$, and AB magnitudes throughout.

\section{Data}
\label{data}
\subsection{Overview}
The data used here  consist of NICMOS parallel observations taken during GO programs 10872 in GOODS and 11236 in COSMOS (PI H. Teplitz), and 11188 in SSA22 (PI B. Siana).   For the GOODS fields, 15 parallel fields were observed in \Jnic\ and \Hnic,  and nine lie  within the GOODS footprint where ACS data are available.  The positions of these fields within GOODS are shown in Figure \ref{goods_footprint} and coordinates are listed in Table \ref{obstable}.        In total, this corresponds to 5.9 square arcminutes of new NICMOS imaging in GOODS.  We note that three fields  (CDFS-1,-2, and -3)  are also included in the \cite{Bouwens08a} search, where they are found not to contain any $z>7$ candidates.   However, in light of the large discrepancy seen in the same NICMOS data by \cite{richard08} and \cite{Bouwens08b}, we include these fields in our search as a consistency check.      Typical exposures for these NICMOS parallels in GOODS were 8 ks in \Jnic\ and 5 ks in \Hnic. 

The COSMOS parallels consist of twelve fields observed in \Jnic\ and \Hnic \footnotemark[1], eleven of which lie within the Subaru/SuprimeCam images in $B$, $r'$, \ip, and \zp. Seven of these eleven fields are also within the ACS \Iacs\ footprint.  A twelfth parallel field lies in the north-east corner of COSMOS, where the limited SuprimeCam coverage is not sensitive enough to discriminate between $z>7$ galaxies and interlopers with typical galaxy colors.  Therefore, we exclude this field from our survey.  
For the remaining eleven COSMOS fields, although the optical imaging is not as deep as in GOODS, it is adequate to remove interlopers, because, as we will show in \S\ 3.1, no $z>7$ candidates are found in the COSMOS fields.   
In total these eleven NICMOS parallel fields cover 7.2 arcmin$^2$. 
Their locations are shown in Figure \ref{cosmos_footprint} and coordinates are listed in Table \ref{obstable}.  Typical exposures were  6-8 ks, divided between \Jnic\ and \Hnic.  
\footnotetext[1]{These 12 fields are distinct from the 500 orbits of \Hnic\ parallel imaging in COSMOS (Colbert et al.\ 2009, in prep), which cannot be used in the $z>7$ search as they lack the essential \Jnic\ imaging.} 

Lastly, we include two parallel fields in  SSA22, which comprise some of the deepest available NICMOS imaging.  However, at these faint limits, optical data in SSA22 that are deep enough to be useful are limited.   Ground-based optical images are not sensitive enough to detect the faintest sources in the NICMOS images, even if they have typical galaxy SEDs.   The only available observation that can adequately rule out interlopers is an ACS  \Iacs\ image (GO  10405, PI S. Chapman), which covers only SSA22-2.  Because all NICMOS sources are detected in this \Iacs\ image, we know that no candidates are found in this parallel field (see \S \ref{selection}) without considering $z$-band data, so we can include it in our survey volume.  Although SSA22-1 can not be used in the $z\sim7$ search,  we are  able to use both fields for the \Jnic -dropout LBG search, because there are no sources that are red enough in \Jnic\ - \Hnic\  to meet the  $z\sim9$ selection criterion in either SSA22 parallel field. 

With these data, we select $z>7$ candidates as $z-$dropouts and \Jnic-dropouts, using the deep optical
images to reject interlopers.  This will be discussed in detail in \S \ref{selection}.

\subsection{NICMOS Data Reduction}
The NICMOS images were reduced and combined with a combination of custom IDL and Python scripts and available IRAF procedures.  First, images were pedestal-corrected, and the South Atlantic Anomaly (SAA) darks were subtracted for impacted orbits.  Following the SAA correction, the pedestal correction was repeated to improve the subtraction.  Next, the sky frames were made and subtracted using McLeod's NICRED (1997) code, and a static bad pixel mask which included the vignetted rows was  created from these sky frames.  To remove any remaining gradients in the images, we made sky images with each column replaced by its median.  This image was smoothed  by a three-pixel wide boxcar and subtracted from each NICMOS frame.  Then this process was repeated for each row of pixels to remove top-to-bottom gradients.  Next, intermittent bad pixels were identified in each image using the IRAF package crutil.  These masks were combined with the static bad pixel mask,  and finally frames were drizzled \citep{drizzle}, using the parameters recommended in the dither handbook (pixfrac = 0.6, and scale  = 0.5).   Shifts were derived so that the final \Jnic\  and \Hnic\ images are drizzled onto the same frame and are therefore aligned.   The resulting pixels are 0\farcs1, and the zero points that we use are adjusted by -0.16 and -0.04 magnitudes in \Jnic\ and \Hnic, to correct for the non-linearity reported by \cite{deJong}.

Sensitivities were measured by randomly placing 0\farcs6 diameter apertures in the images, rejecting apertures which contained light from objects\footnotemark[2].    
\footnotetext[2]{Apertures containing light from objects were identified in two steps.  First, we fit a Gaussian to the full distribution of aperture fluxes, including those that fell on objects.   Then apertures at more than 1 $\sigma$ were rejected and the distribution was re-fit.  This fit mostly relies on the negative side of the flux per aperture distribution.} 
This procedure is repeated for each NICMOS image,  as exposure times varied.  The 5$\sigma$ limits are 26.0-27.5 in \Jnic\ and 25.9-27.0 in \Hnic, with the faintest limits reached in the small area in SSA22 (see Table \ref{obstable}).   The point-spread-function (PSF) for these NICMOS images was derived by  stacking  several isolated, unsaturated stars.   The resulting PSF has a FWHM $\sim$  0\farcs3 in both bands.  The point source aperture correction for a 0\farcs6 diameter aperture is 0.31 magnitudes.

\subsection{Ancillary Optical Data}

\paragraph{GOODS}
We use the publicly available v2.0 ACS GOODS images in \bgoods, \vgoods, \igoods, and \zacs\  bands\footnotemark[3].  Included in v2.0 is additional data used to search for Type Ia supernovae, which doubles the v1.0 exposure time in the \zacs\ band, and also increases the sensitivity in \igoods. This significantly enhances the sensitivity to galaxies at $z\ga 6-7$, and improves identification of faint interlopers.    

As with the NICMOS images,  a PSF is determined by stacking several point sources found in the ACS images.    We find a FWHM  of $\sim$ 0\farcs1 in \zacs.    Typical 3$\sigma$ limits are 28.7, 28.8, 28.3, in  \bgoods, \vgoods, \igoods, measured in 0\farcs4 diameter apertures.   
As we will describe in \S \ref{phot}, ~\zacs\ magnitudes are measured from 0\farcs6 diameter apertures in images matched to the NICMOS resolution.  For these, the  3$\sigma$ sensitivity is $\sim 27-28$ magnitudes.   Some parallel fields near the edge of the GOODS footprint have reduced sensitivity.    We carefully measure the sensitivity in each of the fields, as our objective is to determine whether each source is detected in  \bgoods, \vgoods, or \igoods.  

\footnotetext[3]{http://archive.stsci.edu/prepds/goods/}

\paragraph{COSMOS} 
The COSMOS data that we use are less homogenous than the GOODS data, consisting of both Subaru/SuprimeCam images at $B$, $r'$, \ip, and \zp, and where available,  ACS \Iacs\ images.  The seeing is 0\farcs8 in $B$, $r'$, and \ip, and 1\farcs2 in the \zp\ images.  Typical 3$\sigma$ limits are 28.4, 27.8,  27.3, and 26.7 at $B$, $r'$, \ip, and \zp\ in 1\farcs2  diameter apertures.  The ACS \Iacs\ images typically reach 27.7 in a 0\farcs4 diameter aperture.   Again, sensitivity varies within the COSMOS area, because some of the NICMOS parallels are in the less well covered edges.   As with the GOODS parallel fields, we measure the noise in each field so that we can accurately determine whether sources are detected in the $B$, $r'$, \ip, or \Iacs\ images.

\paragraph{SSA22} 
As described above, the only optical imaging that we use for the SSA22 parallels is an ACS \Iacs\ image that covers SSA22-2.   We use the ``drz'' image, directly from the archive, which has a 3$\sigma$ sensitivity of 28.3 in a 0\farcs4 diameter aperture.   
 
\begin{figure*}
\plottwo{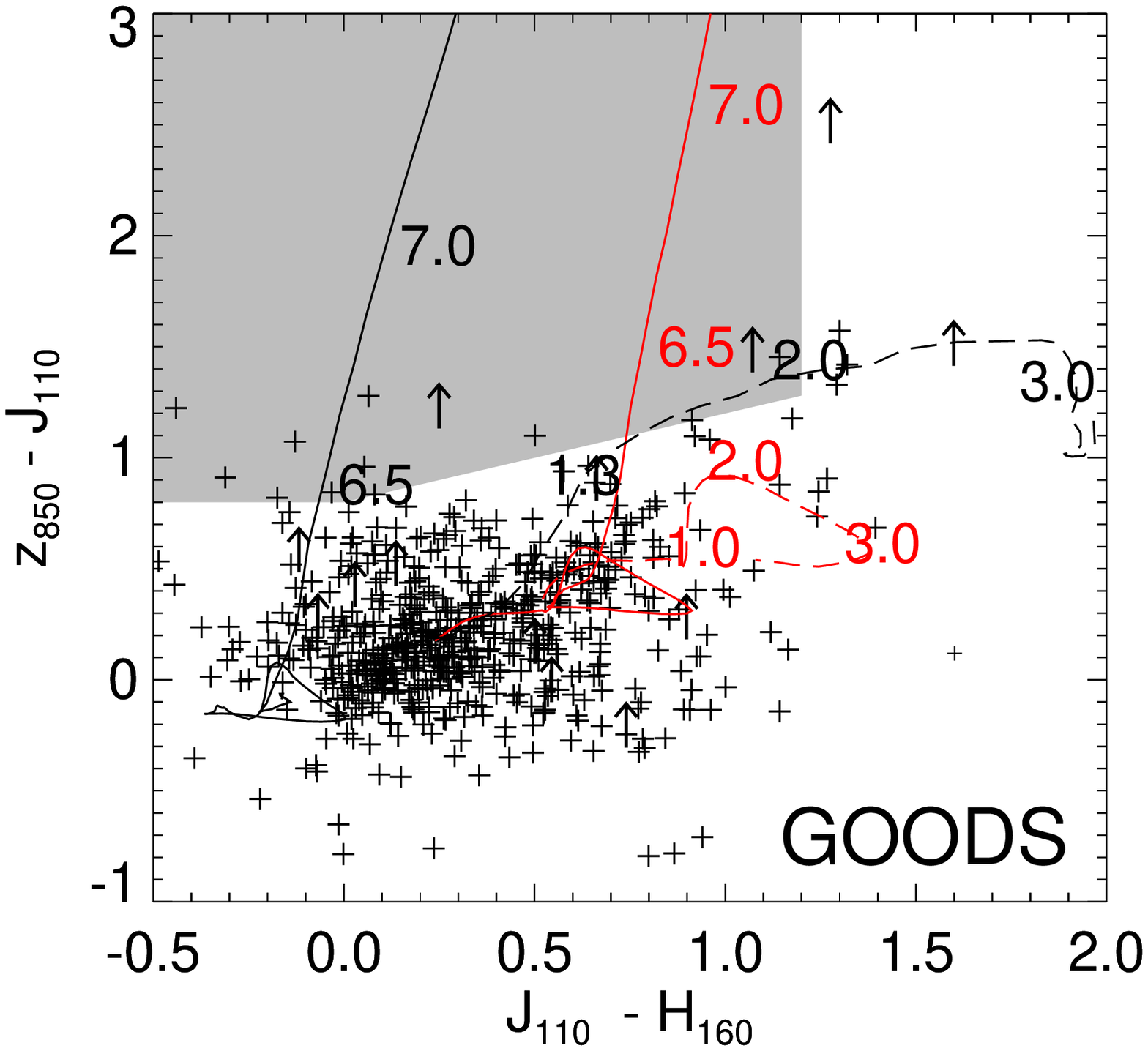}{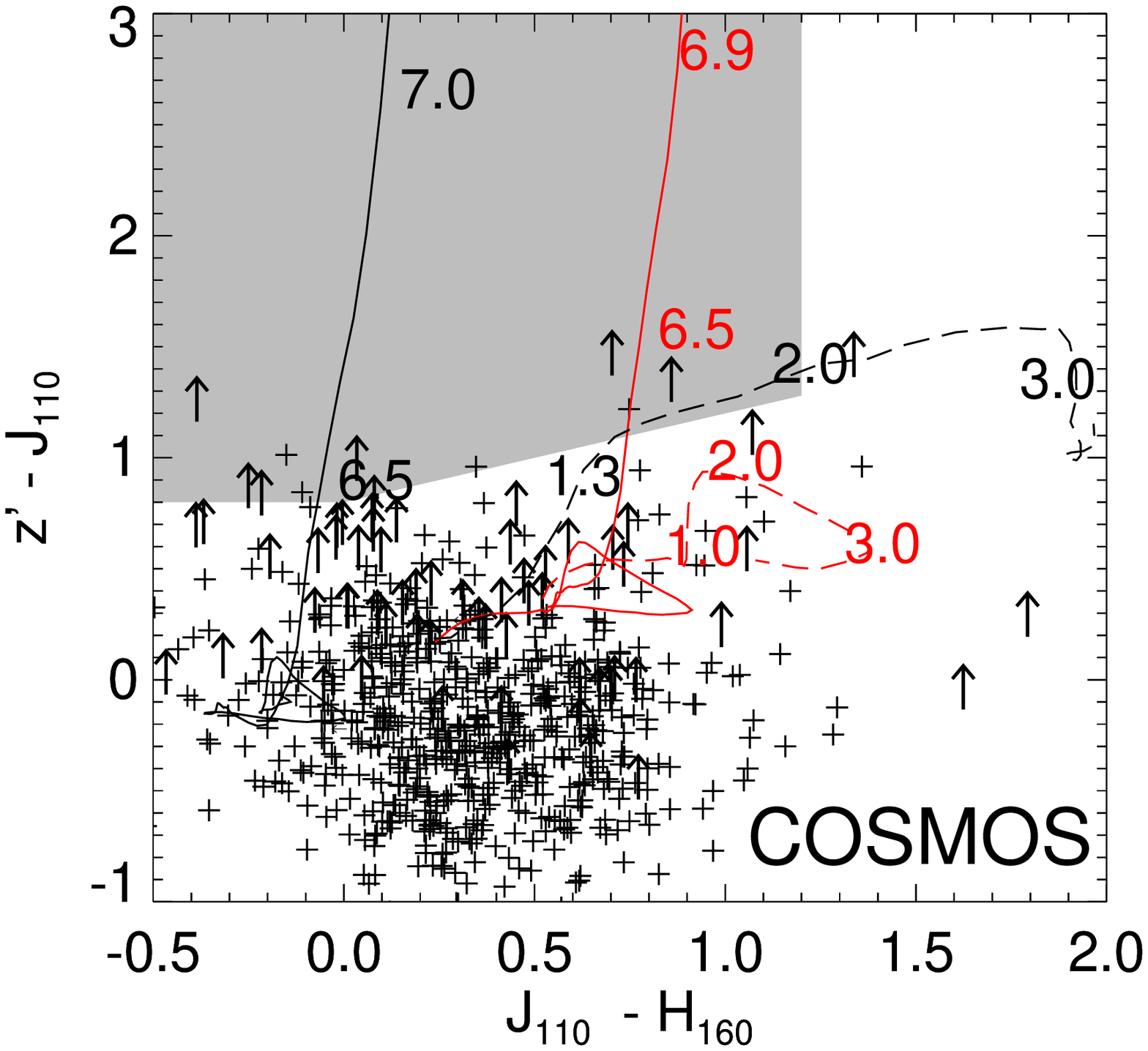}
\caption{The two color plot allows us to identify $z>7$ candidates from the parallels in GOODS (left) and COSMOS (right).   The shaded area shows the color selection adopted by \cite{Bouwens08a}.  Model tracks from \cite{bc03} are: star forming galaxies with E(B-V) =0, 0.5 (solid black, red, respectively), a dusty starburst galaxy (red dashed), and an elliptical (black dashed).  Numbers indicate fiducial redshift points.  }
\label{ccp}
\end{figure*}

\subsection{Photometery} 
\label{phot}

To select $z>7$ galaxies, we compare the above described \zp or \zacs\ data to NICMOS images.   As these data have widely differing resolution,  different techniques are required to measure accurate \zp - \Jnic\ or \zacs\ -\Jnic\ colors.  We describe these approaches below.  

\paragraph{GOODS}

To measure accurate colors of all the galaxies in the nine NICMOS fields in GOODs, we downgraded the resolution of the ACS \zacs\ images by matching the NICMOS PSF.  To achieve this, we use the IRAF task PSFMATCH, which convolves the ACS images with a kernel made from the NICMOS and ACS PSFs.  
The convolved ACS images are then rebinned and aligned with the NICMOS images.  Then, 
we use SExtractor \citep{sextractor}  in dual-image mode, with an inverse-variance weighted average \Jnic\ + \Hnic\ image as the detection image.   For detection, we require five contiguous pixels 1.3 $\sigma$ above the background.   In addition, we use the gauss\_3.0\_5x5.conv filter, which is optimized for finding faint sources.   Lastly, spurious detections, artifacts, and electronic ghosts are manually removed from the catalog.  The \zacs\ - \Jnic\  and \Jnic\ - \Hnic\ colors are measured in 0\farcs6 diameter apertures, and the two-color plot is shown in Figure \ref{ccp} (left).  

In order to test for non-detections at bands shorter than \zacs,  we measure the flux in 0\farcs4  apertures in the original, unconvolved \bgoods, \vgoods,  and \igoods\ images, at the positions predicted by our NICMOS detections.

\paragraph{COSMOS}
 The COSMOS data require a different approach, because downgrading the resolution of the NICMOS 
 images to 1\farcs2 seeing causes a significant loss of sensitivity.  Instead, we resample the \zp\ images to 0\farcs1 per pixel (the same as NICMOS), and align them to match NICMOS.  We then used SExtractor in the same manner as with GOODS, except we use 0\farcs6 diameter apertures in \Jnic\ and \Hnic, and 1\farcs2 in \zp.  The aperture corrections for point sources in these apertures are 0.31 magnitudes for NICMOS and 0.74 magnitudes for COSMOS.   Because galaxies at $z\ga7$ should be compact in NICMOS images (\citealt{Bouwens04_size}; \citealt{Ferguson}; \citealt{DH07}),  this treatment is  appropriate for the sources that we are interested in.  For extended sources, we expect blue-ward scatter in \zp\ - \Jnic\ (away from the $z>7$ selection), as more light will be missed from the higher resolution data.  This trend is confirmed for simulated galaxies, using the IRAF artdata package.   The two-color plot for the COSMOS fields is shown in Figure \ref{ccp} (right).  
 
As with the GOODS data, we measure the flux at the positions predicted by the NICMOS detections, using 0\farcs8 apertures in B, $r'$  and \ip, and 0\farcs4 apertures in \Iacs.

\paragraph{SSA22}
As we are not using any $z-$ band data for SSA22, there is no need to properly account for $z-$ \Jnic\ colors measured with mismatched apertures and resolutions.  Therefore, we simply follow the same procedures described above for  the GOODS and COSMOS parallels-- measuring \Jnic\ and \Hnic\ with SExtractor, and testing for \Iacs\ detections in 0\farcs4 diameters apertures in SSA22-2 where the ACS data are available.

\section{Results}
\label{selection}
\subsection{Selection of $z>7$ candidates}
Candidates for $z>7$ galaxies are selected using the following criteria:
First, we require that galaxies are detected in the \Jnic\ + \Hnic\ detection  image at $>5\sigma$ significance.   In total, we find 696 sources that meet this criterion in GOODS, 701 in COSMOS, and 211 in SSA22.  
Next, $z>7$ candidates must be undetected at the 2 $\sigma$ level in bands bluer than \zp\ or \zacs. This eliminates the vast majority of sources, with only two candidates remaining in the GOODS fields,  two in COSMOS, and none in SSA22-2.  We list these sources in Table \ref{dropouts}, and we will proceed to show that all are interlopers.   

We next use the colors of these sources to determine if any have SEDs consistent with $z\sim7$ galaxies.  We adopt the color cut\footnotemark[4] of Bouwens et al. (2008, 2009), so that candidates must have  $z$ - \Jnic\ > 0.8 and $z$ - \Jnic\  $> 0.8 + 0.4$ (\Jnic - \Hnic), and 
\Jnic\  -\Hnic\ $  < 1.2$ (where $z$ refers to both \zacs\ and \zp).  
   \footnotetext[4]{Despite differing filter set for the COSMOS data, this color cut selects galaxies at $z\ga6.5$ in both cases.  We will show in \S \ref{discussion} that the survey volume is not affected by this inhomogeneity.}   All four of the ``dropout'' sources mentioned above lie outside this selection.  One source (C5-zD1), has   \zp\ - \Jnic\ $\sim$ 0.2, and the others (CDFS3-JD1, CDFS4-JD1 and C8-JD1) have \Jnic\ - \Hnic\ $> 1.2$.  While \cite{Oesch} have suggested a stricter cut of \zacs - \Jnic $> 1.3$, adopting this cut would make no difference in our search, because we have not found any candidates 
   with the most generous selection.  

    The red \Jnic\ - \Hnic\  colors of these three sources could be an indication of the Lyman break in the \Jnic\- band, and redshifts $z>8$ (\citealt{Bouwens05}; \citealt{Henry07}, 2008).  
However,  \Jnic\ - dropouts must also be undetected at the 2$\sigma$ level
 in the \zp\ or \zacs\ bands.   This restriction eliminates  CDFS3-JD1 and CDFS4-JD1.  The remaining source, C8-JD1, can not be 
ruled out on the basis of optical/NIR data alone, but longer wavelength data from IRAC on the {\it Spitzer Space Telescope} show strong detections at 3.6 and 4.5\micron\ (\Hnic\ - [3.6] = 2.4,  \Hnic\ - [4.5] = 2.7).  For $z\sim9$, these colors correspond to a rest-frame UV slope which is much redder than Lyman break galaxies, so this galaxy is more likely an interloper at $z\sim1-3$ with a dusty starburst or an old stellar population.    In conclusion, none of the four optical ``dropout'' sources that we find can be described as a plausible $z>7$ galaxy.  

\subsection{On Incompleteness from  Foreground Contamination}
\label{foreground_contam}
We have rejected as interlopers any sources which have 2$\sigma$ detections in \bgoods, $B$, \vgoods, $r'$, \ip\, or \Iacs.  While this approach is commonly taken in LBG surveys, it does not consider the possibility that a weak detection in any or all of these ``veto'' bands could arise from foreground contamination.  In fact, as we will demonstrate,  the probability of contamination is significant.  

To estimate the influence of foreground contamination, we use the UDF ACS catalogs.  The surface density of sources brighter than our typical 2$\sigma$ detection limits in GOODS (\bgoods, \vgoods\ and  \igoods $\sim$ 29.1, 29.2,  and 28.7) is $\sim$ 400 arcmin$^{-2}$.  
This corresponds to about a 10\% probability of a foreground contaminant lying within 0\farcs5 of a NICMOS detected source.  For the COSMOS fields, the  $B$, $r'$, and \ip\ limits are shallower, so the surface density of possible contaminants is lower ($\sim$ 250 arcmin$^{-2}$).  However, the seeing-limited resolution requires larger apertures.  In this case, we find that the probability of a foreground source lying within 1\arcsec\ of a $z>7$ candidate is about 20\%.  We therefore estimate that 10 - 20 \% of true $z>7$ galaxies would be rejected by our survey because of  faint foreground contaminants.

\section{Discussion} 
\label{discussion}
\subsection{The $z>7$ Luminosity Function} 
While we have not detected any candidate $z>7$ galaxies, we can place limits on the luminosity functions (LFs) of $z-$ dropouts at $z\sim7$, and \Jnic\ -dropouts at $z\sim9$.   Furthermore, we can compare this limit to predictions from LFs at $z\sim3-6$ and place constraints on evolution from $z\sim6$ to 7.  

First, we calculate the survey volume following \cite{Steidel99}:
\begin{equation}
V_{eff}(M) = \int_{z}  {p(M, z) { { dV} \over {dz}} dz}. 
\label{veff_equation}
\end{equation} 
 The quantity $p(M,z)$ is the probability of both detecting a source of a given absolute magnitude and redshift, and selecting it as a $z-$ or \Jnic- dropout based on the criteria that we established in \S \ref{selection}.  This probability can be expressed as 
$p(M, z) = S(M, z) \times C(m) $, with $S(M, z)$ representing the selection function, and $C(m)$  the photometric completeness.   We use simulations to determine these quantities for both $z\sim7$ and $z\sim9$.    To obtain $S(M, z)$, we require only one assumption, namely, a distribution of galaxy spectra.  We use a Gaussian distribution of UV power-law slopes estimated from $z\sim6$ galaxies  ($f_{\lambda} \propto \lambda^{\beta}$ ; $\beta  = -2.2 \pm 0.2$; Stanway et al. 2005).  Then, for every M, $z$, and $\beta$, we predict the \zp, \zacs,  \Jnic\ and \Hnic-- band magnitudes, as well as the the magnitude in the  \Jnic\ + \Hnic\ image that we used for detection.  Sources are required to be (1) bright enough to be detected at $>5  \sigma$ significance in the  \Jnic\ + \Hnic\  image, and  (2) meet the color selection criteria discussed in \S \ref{selection}.  We also assume that 15\% of all $z>7$ galaxies are missed because of foreground contamination, as we showed in \S \ref{foreground_contam}.  We calculate $S(M, z)$ for both the \zp\ (COSMOS) and \zacs\ (GOODS) filter sets, and find that the difference is less than 2\% (for a fixed \Jnic\ + \Hnic\ apparent magnitude limit).  Therefore, the only difference in the GOODS and COSMOS portions of this survey is that the NICMOS images in COSMOS are slightly shallower.  

\begin{figure}
\plotone{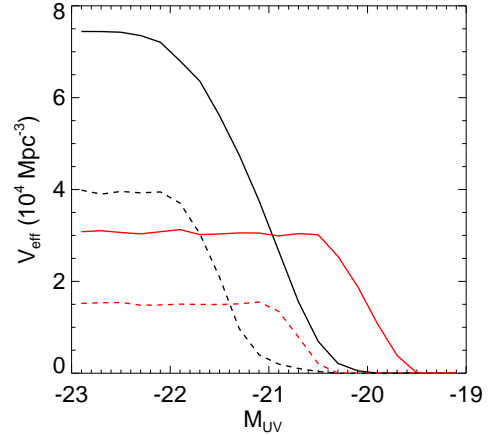}
\caption{The effective volume as a function of absolute UV magnitude for the 14.4 arcmin$^2$ covered by this survey (black), and the 5.8 arcmin$^2$ of the NICMOS UDF (red).  The solid curves are for $z$-band dropouts at $z\sim7$, and the dashed curves are for \Jnic- dropouts at $z\sim9$.   SSA22-1 is excluded from the $z\sim7$ search, as it does not have adequate optical data to rule out interlopers.  The UDF volumes are estimated in the same way as the volume of this survey, assuming a photometric completeness similar to SSA22-1, which has the same sensitivity as the UDF. } 
\label{veff_fig}
\end{figure}

We measure the photometric completeness,  $C(m)$, using the IRAF package, artdata, to add point sources to the 
 \Jnic\ + \Hnic\  images.  We then use SExtractor with the same configuration that we used for the photometry described in \S \ref{phot}.   We find a typical completeness of 70-80\% at the 5$\sigma$ detection threshold for the aggressive SExtractor parameters that we have chosen.  Finally, to evaluate Equation (1), we assume that all of the $z-$dropouts are at $z=7$, and the \Jnic-dropouts are 
 at $z=9$, so that $C(m)$ translates to $C(M)$.   The resulting effective survey volumes for $z\sim7$ and $z\sim9$ are shown in Figure \ref{veff_fig}.     As mentioned in \S \ref{data}, due to limited optical data, we can only include SSA22-2 in the $z\sim7$ search, but both SSA22 fields are included in the $z\sim9$ upper limit, as they contain no candidates with \Jnic - \Hnic\ $>1.2$.

\begin{figure}
\plotone{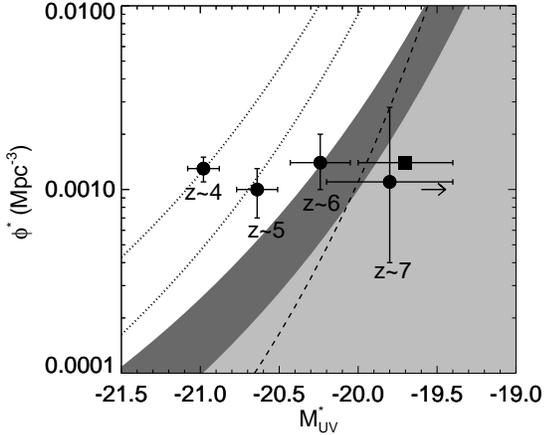}
\caption{Our non-detection of $z>7$ galaxies constrains the luminosity function of these galaxies.   The shaded areas indicate the allowed area for $M^*_{UV}$ and $\phi^*$, for 68\% (light grey) and 95\% (dark grey) confidence.   At $z\sim9$, these same upper limits are shown by the dotted lines, with the allowed parameter space down and to the right.     
 Here, we assume a Schecter parameterization of the LF, with a faint end slope of $\alpha\ = -1.74$ \citep{Bouwens07}.  
 The redshift labels refer to the measurements from Bouwens et al. (2007, 2008), marked by the points,  and the arrow indicates their $M^*$ upper limit at $z\sim9$ when $\phi^* = 10^{-3} ~ {\rm Mpc}^{-3}$.   The square is the LF measurement from \cite{Oesch}, which holds $\phi^*$ fixed at 1.4$\times10^{-3}~ {\rm Mpc}^{-3}$.  The dashed line indicates the upper limit (68\%) from \cite{Mannucci}, again, with the allowed parameter space down and to the right.  } 
\label{lf_lim}
\end{figure}

We next constrain the UV luminosity function.  Assuming a Schecter parameterization of the LF, we show the space allowed for $\phi^*$ and $M^*$ in Figure \ref{lf_lim}.  The shaded areas show the upper limits for 68 and 95\% confidence for the $z\sim7$ survey, and the area below and to the right of the dotted lines indicate the same for the $z\sim9$ search\footnotemark[5]. \footnotetext[5]{Uncertainties given here are in the Poisson noise limit, which is the dominant source of uncertainty when the expected density of sources is $< 1$ arcmin$^{-2}$ \citep{TS08}.  Cosmic variance is also greatly reduced because of the large number of independent sight lines that we have searched. } We also plot measured LFs from  Bouwens et al. (2007, 2008) at $z\sim 4, 5, 6, 7$,  and the upper limit at $z\sim9$.    The non-detections that we find in this survey are consistent with the Bouwens et al. measurements, which predict 0.7 $z\sim7$ candidates in our survey, although the error bars on their $z\sim7$ LF are large, due to the small sample.     The dashed line shows the upper limit  at $z\sim7$ from Mannucci et al. (2007).  Their constraint on luminous $M^*$ is stronger than what we have measured here,  due to their wide area survey ($\sim $130 arcmin$^2$).    Our result is also consistent with the constraint reported by \cite{Stanway08b}, where a $z\sim7$ upper limit that is similar to the $z\sim6$ LF is found. 

This limit can be used  to address the controversy over the numbers of strongly lensed $z>7$ galaxies (\citealt{richard08}; \citealt{Bouwens08b}).   These authors have found differing numbers of candidates behind the same lensing clusters, using the same NICMOS data.  Richard et al. find a few times more candidates than are predicted from the small unlensed sample in the field \citep{Bouwens08a}.  
In fact, such a comparison is difficult to make, as the lensed and field surveys observe mostly different ranges of luminosities.  While the unlensed $<J + H>$ apparent magnitudes of the Richard et al.\ sources range from  27-30, the Bouwens et al. field survey  finds sources down to \Hnic\ $\sim 28$.  However, within this one magnitude of overlap, the Richard et al. density agrees more with the Bouwens et al. measurement at $z\sim6$  than at $z\sim7$.   
While our survey probes even brighter magnitudes, we can compare to the Bouwens et al. $z\sim6$ LF.   Assuming no evolution, this LF predicts 3.2 $z\sim7$ galaxies in our survey volume-- a scenario which we can exclude with 97\% confidence (Poisson statistics).  Our result is more consistent with the  $z\sim7$ result from \cite{Bouwens08a}, as shown in Figure  \ref{lf_lim}.

\subsection{Star Formation and Reionization}
\label{reion}
We also constrain the amount of star formation at $z\sim7$ and 9.  To do this, we fix $\phi^*\sim 10^{-3} ~{\rm Mpc}^{-3}~{\rm mag}^{-1}$.   This is supported by luminosity functions that have been measured by many authors (\citealt{Bouwens07}, and references therein), from $z\sim 3 -6$.  While some scatter is present at $z\sim6$, most LFs agree with this value of $\phi^*$ to within a factor of two, so that any evolution of this parameter must be small.  For this choice of $\phi^*$, we find that $M^* \ge -20.0$ at $z\sim7$, and $\ge -20.7$ at $z\sim 9$.    
Assuming a steep faint end slope  of $\alpha = -1.74$ \citep{Bouwens07},  and integrating the LF to zero luminosity, we find a luminosity density of $\rho_L \le 1.5  \times 10^{26}~ {\rm  erg s^{-1} Hz^{-1} Mpc^{-3}} $ at $z\sim7$. This limit is 1.9 times higher at $z\sim9$.   This corresponds to a star formation density  of $\rho_{SFR} \le 0.019 ~ {\rm M_{\sun} yr^{-1}~Mpc^{-3}}$ at $z\sim7$, when the conversion from \cite{Madau_sfr} is used.  It is important to note that this conversion assumes no extinction, solar metallicity, and a Salpeter IMF with $dN/dM \propto M^{-2.3}$ from $M=0.1 - 100~ M_{\sun}$.    While a correction to  a more likely metallicity of $0.2 Z_{\sun}$  is negligible ($< 5$ \%), a shallower IMF slope of -1.7 will decrease the SFR by a factor of 3.2 (calculated from Starburst99; \citealt{sb99}).  

An important question remains whether galaxies at $z\sim 6-7$ are capable of reionizing the neutral hydrogen in the intergalactic medium. This question is difficult to address, as it depends on the duration of the reionization.  A longer reionization will require more ionizing photons over the lifetime of the galaxies in order to account for recombination \citep{Chary}.   Nonetheless, it is interesting to compare our upper limit to the recombination rate at $z\sim7$, for a completely ionized IGM (consistent with the WMAP 5 year electron scattering optical depth; \citealt{Dunkley}).   \cite{Madau_reion} report this rate in terms of the critical SFR required to maintain an ionized IGM:

\begin{equation}
\rho_{SFR, crit} =
\frac{0.039 ~{\rm  M_{\sun} ~ yr^{-1}~ Mpc^{-3} }}   {f_{esc}}
\left( \frac{1+z}{8}\right)^3
\left(\frac{C}{30}\right)
\left(\frac{\Omega_b h^2}{0.0227}\right)^2,  
\end{equation} 

\noindent where, again,  solar metallicity and a Salpeter IMF from 0.1-100 $M_{\sun}$ are assumed.  This critical SFR also depends on a number of other important, but uncertain parameters.  The escape fraction of ionizing photons, $f_{esc}$, has been difficult to measure.    While a number of authors have found that the escape fraction is small ($<$ 5-10\% relative to photons escaping at 1500\AA\footnotemark[6];  \citealt{Malkan}; \citealt{Siana}, Bridge et al., in prep), there remains some evidence that it could increase with redshift (\citealt{Steidel01}; \citealt{Shapley06}; \citealt{Iwata}).  The HII clumping factor, $C = \langle n_{H II}^2\rangle / \langle n_{H II} \rangle^2$, is also important, as this dictates the average recombination rate per hydrogen atom relative to an IGM of uniform density.  While many authors have adopted an estimate of $C = 30$, based on simulations by \cite{go97}, more recent work suggests that this estimate is much too high, and $C \la 10$ may be more appropriate (e.g. \citealt{BH07}; \citealt{TC07}).  
\footnotetext[6]{These escape fraction upper limits from the literature are the relative escape fraction, described by \cite{Shapley06} and \cite{Siana}, as opposed to the absolute escape fraction that we use in this paper.  By definition, the absolute escape fraction is smaller than the relative escape fraction.  } 

In order to meet the requirement posed by our upper limit of 0.019 ${\rm M_{\sun} yr^{-1}~Mpc^{-3}}$ at $z\sim7$, we find that  $C/f_{esc} \le15$.  However, this number is strongly influenced by the faint end slope of the LF, because we have integrated our constraints to zero luminosity.  We have assumed  a faint end slope of $\alpha = -1.74$, based on \cite{Bouwens07},  but \cite{Oesch07} show that this slope is influenced by input assumptions such as dust extinction and IGM neutral hydrogen absorption (which alter the effective survey volume).  For the shallower slope of $\alpha = -1.6$, reported by Oesch et al., our upper limit is reduced by a factor of 1.7, and we then require $C/f_{esc} \la 9$.  On the other hand, it has been predicted that $\alpha$ approaches --2 for a sample of young galaxies undergoing their first significant bursts of star formation  \citep{Overzier}.    In this case, constraints are more dependent on the true low luminosity cutoff.

  \begin{figure*}
\plotone{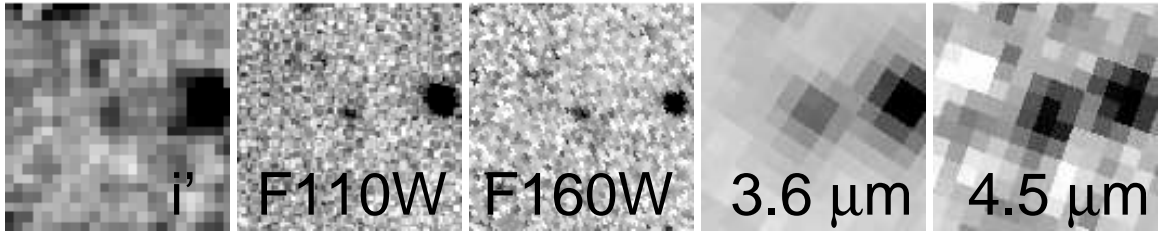}
\caption{Postage stamp images of \jdtwo, from left to right: \ip (Megacam), \Jnic, \Hnic\ (NICMOS), 3.6\micron, and 4.5\micron\ (IRAC).   Images are 7 \arcsec on a side, and are oriented with north up,  and east to the left.  Exposure times in \Jnic\ and 3.6\micron\ are several times longer than their \Hnic\ and 4.5\micron\ counterparts, so the photometry does indeed suggest two spectral breaks. The NICMOS and IRAC images are described in more detail in \cite{Henry08}. }
\label{stampfig}
\end{figure*}

The effects of metallicity and IMF are also important in determining the ionizing output of galaxies.  We use Starburst99 models \citep{sb99} to calculate the ionizing photon rate for metal poor stellar populations ($Z = 0.2 Z_{\sun}$) and for a shallower IMF slope.  For a Salpeter IMF and $Z  = 0.2Z_{\sun}$, a stellar population will produce 1.4 times more ionizing photons than a solar metallicity population with the same UV luminosity.  Consequently, the constraint from this survey becomes $C/f_{esc} < 21$.  Likewise, with $Z  = 0.2Z_{\sun}$ and a shallower IMF slope of  $dN/dM \propto M^{-1.7}$, this constraint is relaxed to $C/f_{esc} < 36$.  

Lastly, it has also been noted that the electron temperature in the primordial HII regions will play an imortant role (e.g. \citealt{Tumlinson};  \citealt{Stiavelli}). Because the recombination coefficient is proportional to $T^{-0.7}$, a factor of two increase in temperature decreases the critical star formation rate by a factor of $\sim 1.6$.  

In summary, we find that for reasonable models,  $C/f_{esc} \la 30-40$ is required to maintain an ionized IGM at $z\sim7$. 
This echos constraints reported by \cite{Chary}, who finds that for $C/ f_{esc} \sim 60$ (``high-V'' case) and a Salpeter IMF the number of ionizing photons produced is too low to reionize hydrogen, unless the reionization occurred rapidly between $6 < z<7$.

\section{Followup of the $z\sim9$ candidate \jdtwo }

\begin{figure*}
\plottwo{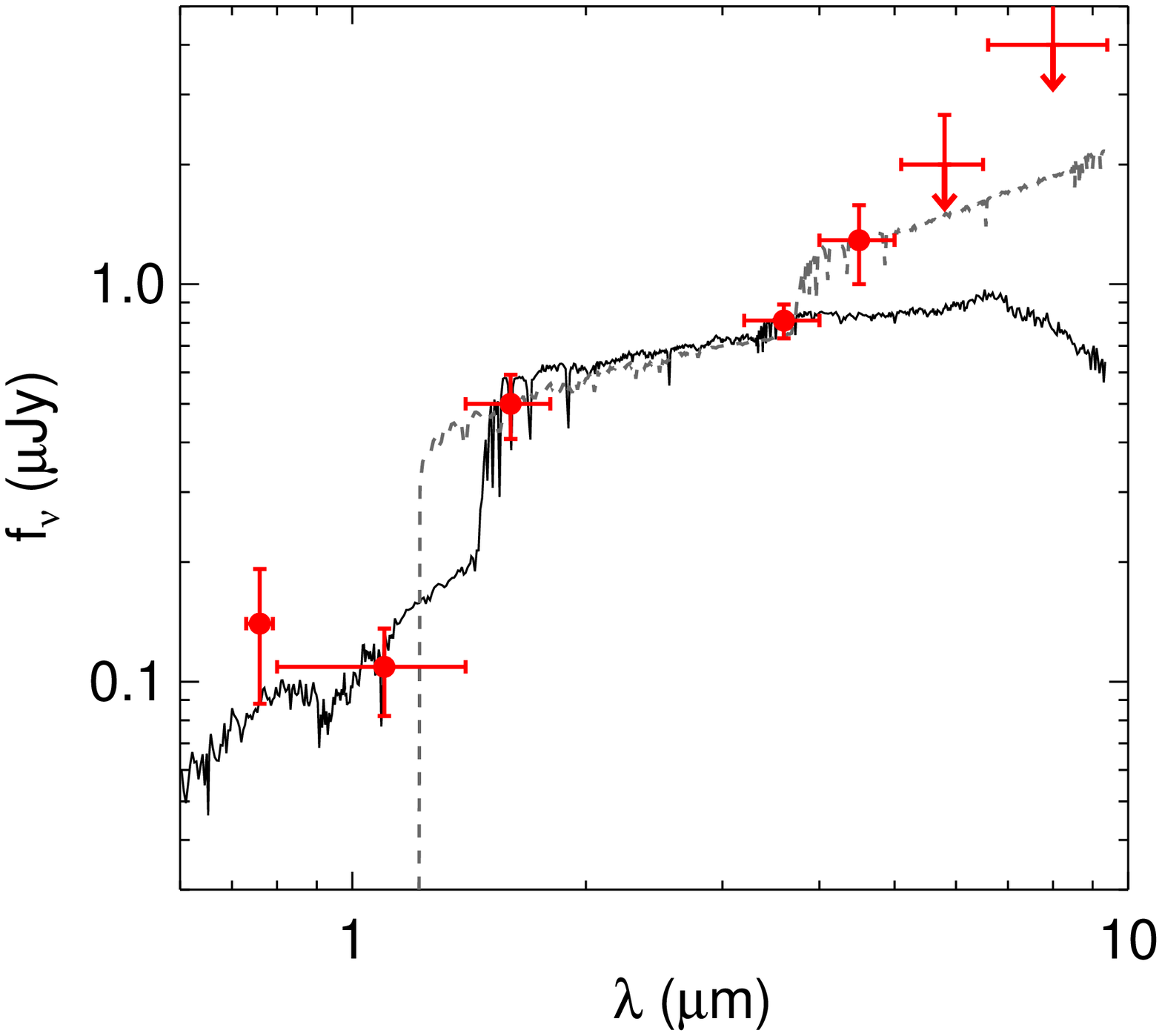}{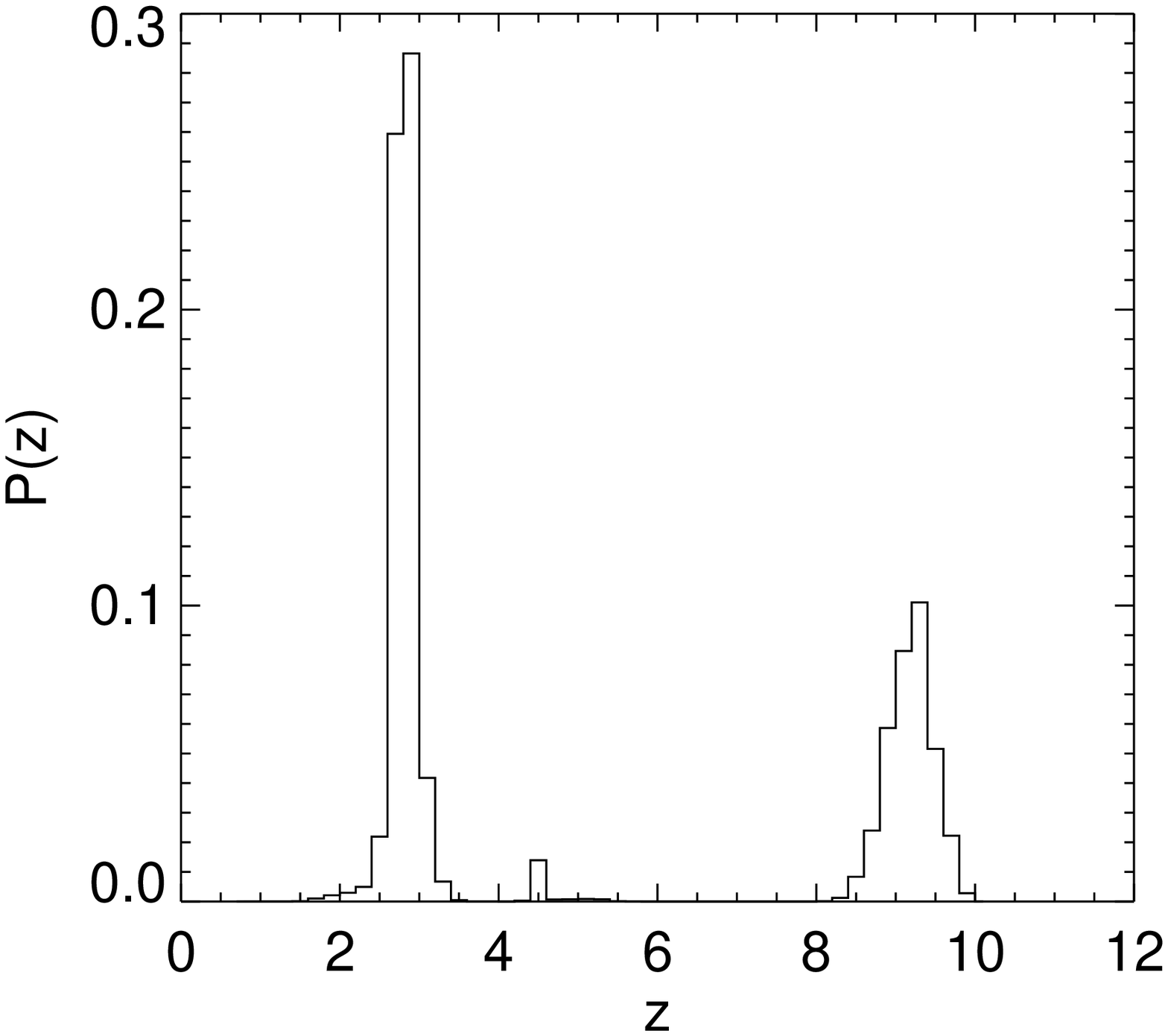}
\caption{ \textit{Left--} The addition of \ip\ improves our photometric redshift, and $z\sim2.7$ is now favored. The solid line is the preferred fit, which is a 250 Myr old instantaneous burst model, with $A_V = 0.2$ and solar metallicity.  The grey dashed line is the best fit $z\sim9$ SED from \cite{Henry08}, which is a 64 Myr old constant star forming model with $A_V = 1.0$ and $Z = 0.2 Z_{\sun}$.  \textit{Right--} The redshift probability distribution from our Monte Carlo simulation shows a peak at $z\sim3$ when the newly acquired \ip\ detection is included. }
\label{newsedfig}
\end{figure*}

\label{jd2_followup}
In \cite{Henry08} we reported the discovery of a luminous $z\sim9$ candidate from the wide area, NICMOS Pure Parallel Survey  (135 arcmin$^2$ to \Jnic\ and \Hnic\ $\sim 25$ AB;  \citealt{Teplitz}; \citealt{Yan00}; \citealt{Colbert}; \citealt{Henry07}).  This candidate, \jdtwo,  was identified as having a strong spectral break between the \Jnic\ and \Hnic\ bands, with a faint but detected \Jnic\ flux and \Jnic\ - \Hnic\ = 1.7.  Subsequent followup observations with {\it Spitzer}/IRAC showed a flat spectrum in \Hnic\ - [3.6], and a second spectral break between 3.6 \micron\ and  4.5\micron.  The only possibility for two breaks are the Lyman and Balmer breaks, and a redshift of $z\sim9$.  However, given the uncertainties in IRAC flux, the significance of the second break is only about 95\%, and without this break, the galaxy spectrum could also be fit by an intermediate-redshift elliptical or post-starburst galaxy.

The main impediment to a robust identification of \jdtwo\ as  a $z\sim9$ galaxy is the lack of deep optical imaging to verify that we have indeed identified the Lyman break.   Such observations require a significant investment, and a non-detection at $I\sim 28$ AB would ultimately not be definitive because interlopers could be even fainter than this.  On the other hand, obtaining a detection would definitively rule out the  $z\sim9$ interpretation.  Therefore, we have obtained \ip\ observations with the MMT to attempt to understand the nature of \jdtwo.

\subsection{MegaCam Observations of \jdtwo}
  The \ip\  observations consisted of a series of exposures, of length 300 to 500\,s each ($\sim6.8$ hours),  taken on the nights of 2008 June 19-24 with Megacam at the 6.5\,m MMT (Mcleod et al.\ 2006).  The observations were carried out through thin cirrus, except for the nights of 2008 June 20 and 21, which were photometric. Seeing varied from as low as 0\farcs8 to as high as 1\farcs6, and averaged about 1\farcs0.  The data were reduced interactively using standard techniques: bias-subtracted and flattened exposures were treated to remove cosmic rays and bad pixels before calculating the coordinates using stars from the USNOB1.0 catalog and correcting the photometry for off-axis scattered light.  The resulting exposures were spatially registered to a common coordinate system.  All frames taken on the nights of 2008 June 19, and 22-24 were then flux-calibrated using exposures from the photometric nights, and all the frames were then coadded to create an \ip\ mosaic. 
  The final image has a seeing FWHM of 1\farcs2. 
  We show a cutout image centered on  \jdtwo\ in Figure \ref{stampfig}, alongside our NICMOS and IRAC images 
  that are described in \cite{Henry08}.

\subsection{Photometry} 
We use the NICMOS images to predict the position of \jdtwo\ in the \ip\ image, and measure the flux in a 1\farcs3 diameter aperture at this position.   The noise is measured by randomly placing apertures in blank parts of the image, as  was done with the NICMOS and other optical images (see \S \ref{data}).  We find a S/N of 2.6, and an aperture magnitude of $26.8 \pm 0.4$. The aperture correction measured for point sources in the field is $2.18 \pm 0.04$ in flux units, and so the result is \ip = $26.0 \pm 0.4$, total.  Although the detection is weak, it strongly suggests an intermediate redshift interloper.  The probability of the \ip\ detection being the result of a foreground contaminant (as we described in \S \ref{foreground_contam}) within 1\arcsec\ of \jdtwo\ is low ($\sim$ 5\%), as the \ip\ image is not as deep as the GOODS and COSMOS optical images.

\subsection{An Updated Photo-z of \jdtwo} 
We update the photometric redshift of \jdtwo\ by including the \ip\ measurement, and repeating the fit that we performed in \cite{Henry08}.  To do this, we use the photometric redshift code,~  {\it Hyperz} \citep{Bolzonella}, with \cite{bc03} stellar synthesis templates.   We fit for redshift, allowing age, extinction (using Calzetti et al.\ 2000), and metallicity to be free ($Z = 0.02, 0.2, 0.4, ~{\rm and} ~1 \times Z_{\sun}$), and using four star formation histories: an instantaneous burst, a constant SFR, and two exponentially declining star formation histories with e-folding times ($\tau_{SFR}$) of 100 and 500 Myrs.  As in \cite{Henry08}, we do not include the upper limits at 5.8 and 8.0 \micron, as they do not constrain the fit.    The revised, best-fitting model is shown in Figure \ref{newsedfig}.   It is described by a 250 Myr  instantaneous burst at $z\sim2.7$, with solar metallicity,  $A_V = 0.2$, and a stellar mass of $9.9 \times 10^9~{\rm M_{\sun}}$.  The absolute B-band magnitude is $M_B = -21.0$.  

We use Monte Carlo simulations to assess this underconstrained  problem by constructing a five dimensional ($z$, age, $A_V$, metallicity, and star formation history) probability density function.  This is done by generating $10^5$ realizations of the photometry, with magnitudes simultaneously perturbed according to the uncertainties.  We then repeat the fit described above.   The probability distribution in redshift space is shown in Figure \ref{newsedfig}.    Now, $z\sim2-3$ solutions  are favored, with 74\% of realizations having a best fit at $z<5$.    The fact that the $z\sim8-10$ interpretation still comprises a significant fraction of the simulated fits is guaranteed by the low S/N \ip\ detection, which frequently dips below 1$\sigma$ when perturbed in the Monte Carlo simulation.  For these cases we do not include the \ip\ observations and the fits strongly favor the $z\sim9$ solution.  Regardless, the inclusion of this weak detection in our analysis adjusts the preferred redshift to $z\sim2-3$.   This is more in line with an extrapolation of the Bouwens et al. (2006, 2007) luminosity functions, which imply a low likelihood of a galaxy at $z\sim9$.  

The additional constraints from our Monte Carlo simulation suggest, for $z<5$: (1) a poorly constrained age with a median of  360 Myrs, and a 68\% confidence interval ranging from 100 Myrs to 1 Gyr and (2) little or no extinction, with  68\% of realizations preferring $A_V$ of 0.5 or less.

\subsection{Interlopers in future $z\sim9$ surveys} 
The discovery that \jdtwo\ is an interloper has important implications for future $z\sim9$ surveys, because similar sources will be readily discovered with new near-infrared instruments.  In addition to \jdtwo\, in the NICMOS Pure Parallel Survey we find 12 more galaxies  down to \Hnic\ $\sim 24$  which have similarly red \Jnic\ - \Hnic $>1.7$.  As this wide area survey is complete for such red galaxies at this limit, the density of these objects is approximately 200 degree$^{-2}$.  
   Longer wavelength IRAC observations of a few of these sources indicate rising SEDs that are indicative of interlopers, but not all of these unusually red galaxies have yet been observed with IRAC.  So it is likely that more galaxies with extremely red \Jnic\ - \Hnic\ and a flat spectrum at longer wavelengths have been detected in the NICMOS pure parallel imaging.   These sources will be difficult, if not impossible to distinguish from $z>7-8$ galaxies in future surveys, meaning that deep optical imaging or a high S/N detection of the Balmer break will be crucial.  

Using deep optical imaging to distinguish $z>7-8$ galaxies from interlopers will be challenging.    At the  faint magnitudes where these galaxies are more likely to be confirmed ($H > 28$ AB), optical observations with the {\it James Webb Space Telescope} will take  at least 10 hours per pointing to reach $\ga 30$ AB at the 2 $\sigma$ level required for non-detection.   In total, this investment in telescope time simply to confirm non-detections could amount to hundreds of hours.    
In addition, as we showed in \S \ref{foreground_contam}, foreground contamination from faint, lower-redshift objects can be a substantial source of incompleteness.  Extrapolating number counts from the UDF, we estimate the surface density of galaxies brighter than $30$ AB  (total magnitudes) in \bgoods, \vgoods, and \igoods\ is $\sim 900$ arcmin$^{-2}$, or 0.25 arcsec$^{-2}$.  Clearly, high angular resolution will be necessary to distinguish interlopers from $z>8$ galaxies, as ground-based seeing limited observations would suffer from severe confusion.

\section{Conclusions} 
The absence of any $z>7$ galaxies in our new NICMOS data strongly constrains the volume density of $z>7$
galaxies.     We have shown that at $z\sim7$, if $\phi^* = 10^{-3} ~{\rm Mpc}^{-3}$, then $M^*_{UV} > -20.0$, and the cosmic star formation density (integrated to zero luminosity) is $<0.019~{\rm M_{\sun}~yr^{-1}~Mpc^{-3}}$.  Although the luminosities that we observe are much brighter than the candidates reported from lensing surveys (Richard et al. 2006, 2008), we can indirectly address their discrepancy with the field survey  of \cite{Bouwens08a}.  Our  non-detection is consistent with Bouwens et al., so our independent result supports their reported evolution for the most luminous sources.  This suggests an additional fading of  $M^*_{UV}$ by 0.4 magnitudes at from $z\sim6$ to $z\sim7$.

Clearly, large uncertainties remain as the few reported candidates are hardly robust detections.    Upcoming surveys using the Wide Field Camera 3 (WFC3) on board HST will address this issue with its improved resolution and sensitivity, increasing the number of known $z\sim7$ candidates by an order of magnitude.  Current plans to use pure parallel mode observations to cover a wide area (PIs M. Trenti,  H. Yan, and M. Malkan) will also provide crucial measurements of the luminous sources.

Interpretation of the UV luminosity function in terms of the ionizing photon budget required for neutral hydrogen reionization 
is uncertain, for reasons that we (in \S \ref{discussion}) and many others (e.g. \citealt{bunker}; \citealt{Bouwens07}) have discussed.  However,  for a  Salpeter IMF and a faint end slope of $\alpha = -1.74$ (reported at $z\sim6$ by \citealt{Bouwens07}), we find  $C/f_{esc} < 15$ is required to maintain a completely ionized IGM at $z\sim7$.  For current estimate of  $C\sim10$ (\citealt{BH07}; \citealt{TC07}) and the commonly adopted $f_{esc} = 0.1$ (e.g. \citealt{Chary}),  this ratio is $C/f_{esc} =  100$--  far too high for star forming galaxies to maintain a completely ionized IGM at $z\sim7$.  However, what is more likely is that  our result provides indirect evidence for significant evolution in one or both of $C$ and $f_{esc}$.  

We also present followup observations of the $z\sim9$ candidate reported in \cite{Henry08}.  With deep imaging from the MMT we find a 2.6$\sigma$ detection at  \ip, which suggests an intermediate-redshift interloper.  This interpretation of the former $z\sim9$ candididate, \jdtwo, is more consistent with upper limits reported by Bouwens et al. (2005, 2008,2009), as well as the upper limit which we find in this study.  The fact that this interloper has such an extremely red \Jnic\ - \Hnic\, and LBG-like SED at longer wavelengths means that similar sources at fainter magnitudes will require a large investment in optical imaging in future surveys, such as those with WFC3 and in the longer term, JWST and future thirty-meter class telescopes.

\begin{deluxetable*}{cccccccccc}
\tablecolumns{10}
\tablecaption{GOODS Fields}
\tablehead{ 
\colhead{ ID}   &  \colhead{RA (J2000)}  & \colhead{Dec(J2000)} & \colhead{\bgoods \tablenotemark{a}  } & \colhead{ \vgoods \tablenotemark{a}}  & \colhead{\igoods \tablenotemark{a}}  &
\colhead{\Iacs\ } & \colhead{\zacs \tablenotemark{b} }    & \colhead{\Jnic \tablenotemark{c} } & \colhead{\Hnic \tablenotemark{c}} 
  }
 \startdata
CDFS-1        & 03 32 26.78 & -27 41 58.4  & 28.7 & 28.9 & 28.3    & \nodata & 28.0  & 26.7 & 26.4 \\
CDFS-2        & 03 32 40.08 & -27 44 03.1  & 28.6 & 28.8 & 28.3    & \nodata & 27.8 &  26.7 & 26.4  \\
CDFS-3        & 03 32 24.31 & -27 40 23.3  & 28.7 & 28.9 & 28.3    & \nodata & 28.0  & 26.7 & 26.4 \\
CDFS-4        & 03 32 52.30 & -27 46 50.9  & 28.7 & 28.5 & 28.2    & \nodata &27.7 & 26.5  & 26.3  \\ 
CDFS-5        & 03 32 27.92 & -27 40 14.8  & 28.8 & 28.8 & 28.2    & \nodata & 27.9 &  26.5 & 25.9  \\
HDFN-1        & 12 37 24.56 & 62 16 23.9   & 28.6 & 28.8 & 28.4    & \nodata &28.0  & 26.5   & 26.1  \\
HDFN-2        & 12 36 06.10 & 62 12 16.1   & 28.7 & 28.9 & 28.4    & \nodata & 28.0 &  26.4 & 26.2 \\
HDFN-3        & 12 27 11.00 & 62 15 57.9   & 28.6 & 28.9 & 28.4    & \nodata & 28.0 &  26.5 & 26.4 \\
HDFN-4        & 12 36 09.09 & 62 06 34.1   &\nodata & 28.5 & 27.9    & \nodata &  27.7  & 26.5  & 26.4 \\
\cutinhead{COSMOS Fields}
\cline{1-10}\\
ID       	   &  RA (J2000)   & Dec  (J2000) & B\tablenotemark{d}  & $r'$ \tablenotemark{d} & \ip \tablenotemark{d} & \Iacs \tablenotemark{a} & \zp \tablenotemark{e} & \Jnic 
\tablenotemark{c} & \Hnic \tablenotemark{c} \\
\cline{1-10} \\ 
COSMOS-1  & 10 01 58.51 & 02 09 35.6   & 28.5 &  27.8 & 27.3    & 27.8      &  26.7  & 26.4  & 26.0  \\
COSMOS-2  & 10 00 32.63 & 01 59 23.0   & 28.8 & 28.1 & 27.7    & 27.3      &  26.9 &26.3   & 26.2  \\
COSMOS-3  & 10 00 30.00 & 02 02 00.0   & 28.8 & 28.1 & 27.7    & 27.8      & 26.8  & 26.0  &  26.2 \\
COSMOS-4  & 10 01 47.25 & 02 56 48.2   & 28.0 & 26.9 & 26.7    & \nodata & 26.0  &  26.1  &  26.2   \\
COSMOS-5  & 10 02 24.91 & 02 51 46.3   & 28.5 & 27.8 & 27.3    & \nodata &  26.7  & 26.0   &  26.1  \\
COSMOS-6  & 09 58 32.62 & 01 48 24.0   & 28.5 & 27.8 & 27.2    & 27.5       & 26.6  &  26.2  &  26.3 \\
COSMOS-7  & 09 58 27.51 & 02 18 29.1   & 28.4 & 27.8 & 27.3    & 26.6       & 26.8  & 26.5  & 26.4   \\
COSMOS-8  & 09 58 40.25 & 02 52 52.7   & 28.4 & 27.8 & 27.3    & \nodata &  26.6 & 26.1  & 26.0 \\ 
COSMOS-9  & 10 02 10.58 & 01 45 46.1   & 28.4 & 27.9 & 27.4    & 27.7       &  26.7 &26.5   &26. 0   \\
COSMOS-10& 09 58 22.61 & 02 39 02.3   & 28.4 & 27.9 & 27.3    & 27.7       &   26.6 & 26.1   &26.3   \\
COSMOS-11& 10 00 02.82 & 02 46 07.8   & 28.4 & 27.8 & 27.3    & \nodata &  26.6 &  26.0   & 26.2   \\
\cutinhead{SSA 22 Fields} 
\cline{1-10} \\ 
ID       	   &  RA (J2000)   & Dec  (J2000) &  &  &  & \Iacs \tablenotemark{a} & & \Jnic \tablenotemark{c} & \Hnic \tablenotemark{c} \\
\cline{1-10} \\
SSA22-1        & 22 17 21.23  & 00 24 09.8 & &  & & \nodata &  & 27.4 & 27.0 \\ 
SSA22-2       &  22 17 23.36  & 00 22 03.6  & &  &  & 28.3 &  & 27.0 & 26.5 \\
\enddata
\tablecomments{Sensitivities were measured by randomly placing apertures in blank parts of the images.  All limits are in aperture magnitudes, and aperture corrections are given in \S 2. } 
\tablenotetext{a}{3 $\sigma$ limits measured in 0\farcs4 diameter apertures.}
\tablenotetext{b}{3 $\sigma$ limits measured in images that were PSF-convolved to match the NICMOS resolution, using 0\farcs6 diameter apertures.}
\tablenotetext{c}{5 $\sigma$ limits measured in 0\farcs6 diameter apertures.}
\tablenotetext{d}{3 $\sigma$ limits measured in 0\farcs8 diameter apertures.} 
\tablenotetext{e}{3 $\sigma$ limits measured in 1\farcs2 diameter apertures.} 
\label{obstable}
\end{deluxetable*}

\begin{deluxetable*}{cccccc}
\tablecolumns{9}
\tablecaption{Optical Dropout Sources}
\tablehead{ 
\colhead{ ID}   &  \colhead{RA (J2000)}  & \colhead{Dec (J2000)}  & $z -$ \Hnic\ & \Jnic\ - \Hnic\ & \Hnic\   }
\startdata
CDFS3-JD1 &  03 32 23.24  & -27 40 20.8   & 1.7  & 1.6 & 25.1 \\  
CDFS4-JD1 & 	03 32 51.66  & -27 47 15.3   & 1.7    &  1.5      & 25.3 \\
C5-zD1  & 10 02 24.47   &  02 52 05.4  &  0.2  & -0.1 & 25.7 \\ 
C8-JD1   &  09 58 39.07  & 02 52 53.6  & $>1.2$ &  $>1.6$ & 25.1 \\   
\enddata
\tablecomments{\Hnic\ magnitudes are aperture corrected, assuming a point source correction of 0.31 magnitudes.  Here, $z$ refers to \zacs\ for the GOODS sources, and \zp\ for the COSMOS sources.   Non-detections are 2$\sigma$.  } 
\label{dropouts}
\end{deluxetable*}

\acknowledgements
This work is funded in part by the University of California President's Dissertation Year Fellowship. 
Observations reported here were obtained at the MMT Observatory, a joint facility of the Smithsonian Institution and the University of Arizona.   The authors would like to thank S. Furlanetto for helpful discussions, and K. Kornei  and R. Bouwens for comments that improved this manuscript.  

{\it Facilities:} \facility{MMT:Megacam}, \facility{HST}, \facility{Spitzer},

\end{document}